\documentclass[12pt]{article}
\usepackage{epsfig}
\def\be{\begin{equation}}
\def\ee{\end{equation}}
\def\bea{\begin{eqnarray}}
\def\eea{\end{eqnarray}}

\title { \bf Time dependent action in $\phi^6$ potential}
\author{Hatem Widyan\thanks{E--mail : widyan@aabu.edu.jo }\\
    {Department of Physics} \\
    {Al al-Bayt University, Mafraq 25113, Jordan} \\
    and \\
    {Mashhoor Al-Wardat} \\
       { Physics Department} \\
    {Yarmouk University} \\
    { P.O.B. 566 Irbid, 21163 Jordan}
        }

\begin{document}
\maketitle
%
%
%
{\bf Abstract:}  The false vacuum decay in field theory from a
coherently oscillating initial state is studied for $\phi^6$
potential. An oscillating bubble solution is obtained. The
instantaneous bubble nucleation rate is calculated.


{\bf keywords:} phase transition, tunneling, scalar field theory.

{ \bf PACS numbers:} 03.65.Sq, 04.62.1v

%
%

\begin {section}{\bf Introduction}
\par

Problems involving quantum mechanical tunneling in a time
dependent setting may arise in a wide variety of contexts, such as
Schwinger vacuum pair production for time-dependent laser pulses
\cite{cesim}, pair creation of charged particles in time dependent
background electromagnetic fields \cite{brezin,popov,audretsch},
quantum interference in vacuum pair production \cite{dunne},
Hawking radiation from black holes \cite{jiang}, spontaneous
nucleation of topological defects in expanding universes
\cite{vijay} and false vacuum decay with time dependent initial
states or time dependent potentials \cite{widrow,esko}.

Barrier penetration and tunneling for a particle moving in a
one-dimensional potential are treated in all textbooks on quantum
mechanics. The procedure is by making a WKB approximation and
expanding the logarithm of the wave function in powers of $\hbar$.
An alternative way to tunneling makes use of the
Euclidean-path-integral (EPI) formulation of the theory
\cite{Coleman}. According to Feynman \cite{feynman}, the amplitude
for going from one state to another is given by the sum over all
paths connecting the states weighted by ${\rm e}^{i S/\hbar}$,
where $S$ is the action evaluated along the path. For classically
allowed motion, the dominant contribution to the path
corresponding to the solution of the real-time equation of motion.
A convenient way to calculate the action is to switch to Euclidean
time. In this case, the probability amplitude is ${\rm e}^{-
S_E/\hbar}$, where $S_E$ is the difference of Euclidean actions
between the instanton solution (instanton solution: the classical
solution of the Euclidean equation of motion with appropriate
boundary conditions) and false vacuum solution.

Most decay of the false-vacuum calculations in single scalar field
theory make use of the EPI formalism. The Lagrangian of the
theory is
 $$
 {\mathcal L} = \frac{1}{2} \partial_\mu \phi \partial_\mu\phi -
 V(\phi),
$$
 where $V(\phi)$ is a potential which has two nondegenerate
 minima: $\phi_+$ ($\phi_-$) is the false (true) vacuum.
One begins by writing the Euclidean action and the equation of
motion. The equations are solved to obtain the instanton solution
with boundary conditions $\phi \rightarrow \phi_+$ for $\eta
\rightarrow \infty$ and $\phi\simeq \phi_-$ for $\eta \rightarrow
0$, where $\eta= \sqrt{\tau^2 + r^2}$ and $\tau$ is the Euclidean
time. The instanton solution corresponds to a bubble being
nucleated at $r=0$.

The bubble nucleation rate per unit time per unit volume is given
by
 $$
 {\Gamma} = A\, {\rm e}^{- S_E/\hbar},
 $$
 where $S_E$ is the difference of Euclidean action  and $A$ is a constant.
 The relevant solution is the one  which
 gives the least action. In flat space and at zero temperature,
 the dominant contribution comes from the unique O(4)-symmetric
 solution \cite{Glaser}.

 As pointed out in \cite{widrow}, EPI has several limitations. We
 are lost at the outset if $\phi$ couples to some external current
 or field which is time dependent. As an example of this case is a
 scalar field in a Friedmann-Robertson-Walker (FRW) cosmology,
 since the FRW space is a time dependent and cannot be written in
 static coordinates. Another example arises with the theories of two
 or more coupled field. Also, there is a limitation of EPI
 formalism in the theory of a single scalar field in flat space if
 the initial field configuration is more complicated than simply
 $\phi(\vec x) = \phi_+$ or time-dependent potential.
 In this work we can consider the case where $\phi$ is
 homogenous and undergoing coherent oscillations about the false vacuum.

One approach to overcome these limitations is presented in
\cite{widrow}. The author studied the false-vacuum decay of a
scalar field by making use of the functional Schrodinger equation. He studied the vacuum decay of a scalar field coupled to a
time-dependent external field and derived the traversal time for
bubble nucleation.

An alternative approach is presented in \cite{esko}. The authors
presented a method  based on WKB approximation combined with complex
time path methods, which can be used to calculate the relevant
tunneling probabilities. They applied their algorithm to
production of charged particle-antiparticle pairs in a
time-dependent electric field and false vacuum decay in field
theory from a coherently oscillating initial state. For the field
theory example, they considered the potential discussed in
Coleman \cite{Coleman},
 $$
   V(\phi) = \frac{\lambda}{2} (\phi^2-a^2)^2+
   \frac{\epsilon}{2a}(\phi-a).
$$
The influence of nontrivial background and decoherence on
vacuum tunneling is presented in \cite{queisser}. In this work we
follow the algorithm presented in \cite{esko}, and we discuss the
effect of coherent oscillating false vacuum state on vacuum decay
in the thin-wall approximation (TWA),
 but we choose the $\phi^6$ potential which was investigated
by many authors in the context of condensed matter as well as
particle physics (see for example
\cite{Bergner,Amaral,Flores,Joy,Arnold,Zamo,Lu,Kim,hatem,hatem1}).
We have noticed that there is a large correction to the nucleation
rate and  the small oscillations about the false vacuum rendered the
state more unstable.

The plan of this paper is as follows. In section $2$, the  vacuum
decay without oscillation about the false in TWA is discussed
using Coleman's approach. In section $3$, decay with oscillation
about the false vacuum in the TWA is presented based on complex
time method. In section $4$ the structure of the oscillating
bubble is obtained, while in section $5$ bubble nucleation decay
rate is calculated. Finally, the results are discussed.

\end{section}

%
%

\begin{section}{\bf Decay without oscillation about the false vacuum: Coleman's
approach}

\indent Let us consider a scalar field theory with a Lagrangian density

$$
 {\mathcal{L}(\phi)}  =  {1 \over
2}~({\partial_{\mu}\phi})^2 - V(\phi), \nonumber
 $$
where the potential $V(\phi)$ is the effective potential at zero
temperature and is given by
\be
 V(\phi) = g \phi^6- 2 g \lambda^2 \phi^4 +(g
 \lambda^4-\delta)\phi^2 , \label{potential}
\ee
we choose $g=0.07$ and $\lambda=2.39$. The potential is shown in Figure 1,
it has two nondegenrate minima
$\phi_+$(false vacuum) and $\phi_-$ (true vacuum) which are all independent of time.

\begin{figure}
\epsfig{file=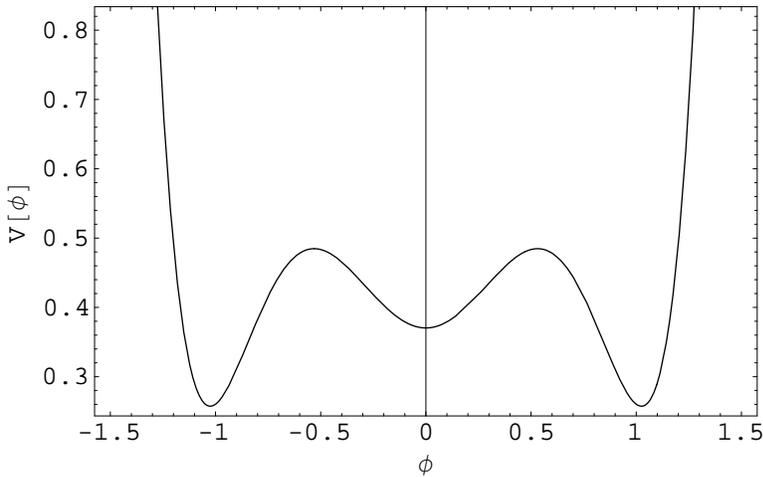}
   \caption{ The scalar field potential $V(\phi)$ with
parameters: $g=0.07$, $\lambda=2.39$, and $\delta=0.2$.}
\end{figure}

Let us expand the true vacuum in powers of $\delta$
$$
 \phi_- = \lambda + e_1 \delta + e_2 \delta^2 + ... \nonumber
$$
To first order in $\delta$,
$$
 \phi_- = \lambda + \frac{1}{4 g \lambda^3} \delta +\mathcal
 O(\delta^2),
$$
$$
 V(\phi_-) = - \delta \lambda^2 + \mathcal O(\delta^2)
4$$
 Similarly for the false vacuum

$$
 \phi_+ =  0+ e_1 \delta + e_2 \delta^2 + ...,
$$
and
$$
 \phi_+ =  0~ {\rm for\,~ all\, ~orders ~\, of\,} ~ \delta,
$$
$$
 V(\phi_+) = 0~{\rm for\, ~all\, ~orders \,~ of\,}  ~\delta
$$

To calculate the probability of decay of the false vacuum in
quantum field theory at zero temperature, one should first solve
the Euclidean equation of motion of the instanton:
\begin{equation}
 \partial_{\mu}\partial_{\mu}\phi = {dV(\phi) \over d\phi}
\label{equa:motion} ,
\end{equation}
with the boundary condition
$\phi \to \phi_+$ as $ \vec x^2+\tau^2 \to \infty$ ,
where $\tau$ is the imaginary time. The probability of tunnelling per unit
time per unit volume is given by
\begin{equation}
\Gamma  =   A ~~  e{^{-S_E[\phi]}},
\end{equation}
where $ S_E[\phi]$ is the Euclidean action corresponding to the solution
of Eq.~(\ref{equa:motion}) and given by the
following expression :
\begin{equation}
S_E[\phi] =  \int d^4{x} \left[ {1 \over 2} \Big({\partial\phi \over
\partial\tau}\Big)^2 + {1 \over 2} \Big(\nabla\phi\Big)^2 +
V(\phi) \right].
\end{equation}
 Since we are interested in the lowest-action instanton, we can reduce
the problem to one of one degree of freedom. If we assume O(4)
rotational symmetry in Euclidean space, then an O(4) invariant
solution of Eq.~(\ref{equa:motion}) exists and its action
$S_E[\phi]$ will be lower than that of any O(4) noninvariant
solution \cite{Glaser}. In this case Eq.~(\ref{equa:motion}) takes
the simpler form
\begin{equation}
{d^2\phi \over d\eta^2} + {3 \over \eta}~ {d\phi \over d\eta} =
{dV(\phi) \over d\phi }, \label{eom00}
\end{equation}
where $\eta=\sqrt{\vec x^2+\tau^2}$, with boundary conditions
$$
 \mathrm{ \phi \to \phi_+  ~~ as} ~~ \eta \to \infty ~ ,~~ {d\phi
\over d\eta }= 0 ~~ \mathrm{ at ~~ \eta = 0 } .
 $$
 We denote the action of this solution by $S_0$. There is an interesting
 case (in the sense that the action can
be calculated analytically) when
\be
 V(\phi_+)-V(\phi_-) =
\lambda^2 \delta +\mathcal O(\delta^2)\equiv \rho_0
\ee
 is much smaller than the
height of the barrier. This is known as the thin-wall
approximation (TWA) and the equation of motion (Eq.~\ref{eom00}) becomes
\be
 \frac{d^2\phi}{d\eta^2}=\frac{dV(\phi)}{d\phi}, \label{soltwa}
\ee
which can be solved analytically for some potentials. For the
$\phi^6$ potential, the solution has the form \cite{Flores,Joy}
\begin{equation}
\phi^2_{\rm wall}(\eta)=\frac{\lambda^2}{1+{\rm e}^{\mu \eta}},
\label{phiwall}
\end{equation}
where $\mu=\sqrt{8 g}\lambda^2$, and $\mu^2$ is the second
derivative of the potential in the TWA limit evaluated at
$\phi_-$.

The action $S_0$ of the O(4)-symmetric bubble is equal to
\begin{eqnarray}
S_0 & = &2 \pi^2 \int_0^\infty d\eta ~ \eta^3 ~\left[ {1 \over
2}\Big({d\phi \over d\eta}\Big)^2 + V(\phi) \right]  \nonumber \\
 & = & -{1 \over 2} \pi^2 \rho_0 R_0^4 + 2 \pi^2 \sigma_0 R_0^3
.
\end{eqnarray}
 Here $R_0$ is the radius of the bubble and $\sigma_0$ is the bubble
 wall surface energy (surface tension), which is given by
\begin{eqnarray}
\sigma_0 & = & \int_0^\infty d\eta \left[ \Big({d\phi \over
d\eta}\Big)^2 + g \,
\phi^2 (\phi^2-\lambda^2)^2 \right] \nonumber \\
& = & - \int_{0}^{\lambda} d\phi \> \sqrt{2 g \, \phi^2
(\phi^2-\lambda^2)^2} \nonumber \\
& = & \frac{\sqrt g \lambda^4}{2 \sqrt 2} \label{10:10},
\end{eqnarray}
and the integral should be calculated in the limit $\rho_0 \to 0$.

The bubble radius $R_0$, is calculated by minimizing $S_0$, this
gives us
 $$
R_0= {{3 \sigma_0} \over \rho_0},
 $$
whence it follows that
\begin{equation}
S_0={{27 \pi^2 \sigma_0^4} \over {2 \rho_0^3}}. \label{S00}
\end{equation}
The nucleation rate is then
\be
 \Gamma  =   A  {\rm e}^{-S_0} = A  {\rm e}^{-\frac{\pi^2}{6}
\rho_0 R_0^4}.
\ee
Another parameter which is defined to test the applicability of
the TWA is the bubble wall thickness $L$ which must be much less
than $R_0$ and is given by
\begin{equation}
L=\frac{1}{\mu}=\Bigg(\frac{d^2 V(\phi_-)}{d\phi^2}\Bigg)^{-1/2},
\label{thickness}
\end{equation}
 The same results can be obtained using the algorithm proposed in
 \cite{esko}.

To summarize, in the TWA the instanton takes the following shape:
\be
 \phi(\eta) = \left\{
  \begin{array}{lll}
    \phi_+ = \lambda +\mathcal O(\delta), & { \eta <<R ~ ({\rm
     True~vacuum})}\\[0.2cm]
   \phi_{\rm wall}(\eta)=\frac{\lambda^2}{1+{\rm e}^{\mu \eta}}, & { \eta \sim
   R}
    \\[0.2cm]
     \phi_- = 0, & {\eta >>R ~ ({\rm False~vacuum})}.
     \end{array}
      \right. \label{phi0}
\ee

\end{section}

%
%

\begin{section}{\bf Decay with oscillation about the false vacuum: Complex
time method}

In this section we review the results obtained in \cite{esko}. We
assume that the field is initially oscillating around the false
vacuum $\phi_f$  and takes the form
\be
 \phi_f(t) = \phi_+ + \alpha_0 \, {\rm sin}\, \omega t.
\ee

Since the energy is conserved, then $E({\rm inside})+ E({\rm
wall})$ of the bubble must equal the energy present in the region
before nucleation of the bubble: $E_{\rm initial}.$ Thus
 $$
 E_{\rm bubble}({\rm inside})= \frac{4\pi}{3} V(\phi_t) R^3,
 $$
where $\phi_t$ is the true vacuum and the bubble wall has the
energy
 $$
 E_{\rm bubble}({\rm wall})= \frac{4\pi \sigma_E^{\rm bubble}
 R^2}{\sqrt{1-\dot  R^2}},
 $$
where
 $$
 \sigma_E^{\rm bubble} = \int_{\rm wall} dr \Bigg[ \frac{1}{2} (\dot
 \phi_{\rm
 bubble})^2
 + \frac{1}{2} ( \phi'_{\rm bubble})^2 + V(\phi_{\rm bubble})
 \Bigg].
 $$

The initial energy from the false vacuum $\phi_f(t)$ has two
contributions, namely
 $$
 E_{\rm initial}({\rm inside}) = \frac{4\pi}{3} R^3 \Bigg[
   \frac{1}{2} (\dot \phi_f(t))^2 + \frac{1}{2} ( \phi'_f(t))^2
   + V(\phi_f(t)) \Bigg] = \frac{4\pi}{3}  \rho_E^{\rm FV} R^3,
 $$
and
 $$
  E_{\rm initial}({\rm wall})= \frac{4\pi \sigma_E^{\rm FV}
  R^2}{\sqrt{1-\dot  R^2}},
 $$
where
 $$
 \sigma_E^{\rm FV} = \int_{\rm wall} dr \Bigg[ \frac{1}{2} (\dot
 \phi_f(t))^2
 + V(\phi_f(t))
 \Bigg].
 $$

From conservation of energy

 $$
 E_{\rm bubble}({\rm inside})+ E_{\rm bubble}({\rm wall}) =
E_{\rm initial}({\rm inside}) + E_{\rm initial}({\rm wall})
 $$
which can be written as
 $$
 \frac{4\pi \sigma_E
  R^2}{\sqrt{1-\dot  R^2}} - \frac{4\pi}{3} \rho_E  R^3=0,
 $$
with
\be
 \sigma_E = \sigma_E^{\rm bubble} - \sigma_E^{\rm FV},
\ee
and
 \be
 \rho_E = \rho_E^{\rm FV} - V(\phi_t).
\ee

From the above two equations, we can define the radius of the
bubble at some time $t_0$ as
\be
 {\mathcal R_0} ={3 \sigma_E \over \rho_E},
 \ee
and the radius at any later time $t$ (the trajectory) is
\be
 R(t) = \sqrt{{\mathcal R_0}^2 + (t-t_0)^2}.
\ee

The action ($S= S_{\rm bubble} - S_{\rm FV}$) is integrated over
an imaginary time contour running from some initial time $t_0$ to
$t_0+i {\mathcal R_0}$, where the bubble shrinks to zero size. The
bubble action is given by
\be
 S_{\rm bubble} = - \int dt \Bigg[4 \pi \sigma_L^{\rm bubble}(t)
  R^2(t) \sqrt{1-\dot R^2}  + \frac{4\pi}{3} V(\phi_t) R^3
 \Bigg], \label{S_buuble}
 \ee
where
\bea
 \sigma_L^{\rm bubble}& = & - \int_{\rm wall} dr \Bigg[\frac{1}{2}
 \Big(\dot \phi_{\rm bubble} \Big)^2 -
 \frac{1}{2} \Big( \phi'_{\rm bubble} \Big)^2 - V(\phi_{\rm bubble})
  \Bigg] \nonumber \\
 & = & \sigma_E^{\rm bubble} - \int_{\rm wall} dr \, \dot\phi^2_{\rm
 bubble} \nonumber
\eea
while the  false vacuum action is
\be
 S_{\rm FV} = - \int dt \Bigg[4 \pi \sigma_L^{\rm FV}(t) R^2(t)
  \sqrt{1-\dot R^2}  + \frac{4\pi}{3} \rho_L^{\rm FV} R^3 \Bigg]
  \label{S_fv},
 \ee
where
 $$
 \sigma_L^{\rm FV} =  \sigma_E^{\rm FV}- \int_{\rm wall} dr \, \dot
 \phi^2_f
 $$
 and
 $$
 \rho_L^{\rm FV} =  \rho_E^{\rm FV}- \dot \phi^2_f
 $$
From Eqs. (\ref{S_buuble}) and (\ref{S_fv}) the action is
\be
 S = - \int dt \Bigg[ 4 \pi \sigma_L(t) R^2(t) \sqrt{1-\dot R^2} -
  \frac{4 \pi}{3} \rho_L(t) R^3
 \Bigg], \label{action}
\ee
where
 $$
   \sigma_L(t) =  \sigma_E - \int_{\rm wall} dr
    \,[\dot \phi^2_{\rm bubble} - \dot \phi^2_f],
 $$
 $$
 \rho_L(t) =  \rho_E - \dot \phi^2_f .
 $$

\end{section}

%
%

\begin{section}{\bf Structure of the Oscillating Bubble }

We calculate the oscillating bubble $\phi_{\rm bubble}(r,t)$ for
the $\phi^6$ potential  which interpolates between the true
 vacuum $\phi_t$ and the false vacuum $\phi_f$. Since the
 initial state oscillates coherently then it breaks the
 symmetry of the theory from SO$(3,1)$ to SO$(3)$. Therefore,
  Eq.~(\ref{equa:motion}) becomes
\be
 \ddot\phi - \frac{1}{r^2} (r^2 \phi')' = - \frac{dV}{d\phi}.
 \label{eom}
\ee
with the potential
 $$
 V(\phi) = g \phi^6- 2 g \lambda^2 \phi^4 +(g
 \lambda^4-\delta)\phi^2.
 $$

Following \cite{esko}, we will find a time-dependent solution
$\phi_{\rm bubble}(r,t),$ which will be reduced to coherently
oscillating field
\be
 \phi_f(t) = \phi_+ + \alpha_0\, {\rm sin}\, \omega t \label{phif}
\ee
about the false vacuum as $r \rightarrow \infty.$ The frequency of
the oscillations $(\omega)$ about the false vacuum is
 $$
 \omega^2 = \frac{d^2V}{d\phi^2}(\phi_+) = 2(g \lambda^4 -
 \delta),
 $$
and its range is $0 < \omega^2 < 4.57.$

We assumed the $\phi_{\rm bubble}(r,t)$ is a function of both space
 and time and takes the form
\be
 \phi_{\rm bubble}(r,t) = \phi_0(r) + \alpha(r) \, {\rm sin}\,
 \omega t.   \label{bubble}
\ee

After substituting Eq.(\ref{bubble}) in Eq.(\ref{eom}) we get
\be
 \alpha''(r) +\frac{2}{r} \alpha'(r) +\left[\omega^2
-\frac{d^2V}{d\phi^2}(\phi_0) \right]
 \alpha(r) =0 . \label{eom2}
\ee
Now we will solve the above equation of motion in three different
regions.

 Firstly, the region outside the bubble $(r>R)$. In
this case, $\phi_0=\phi_+$ (false vacuum),
$\frac{d^2V}{d\phi^2}(\phi_0) \rightarrow
\frac{d^2V}{d\phi^2}(\phi_+) =\omega^2$, and Eq.~(\ref{eom2})
becomes
 $$
 \alpha''(r) +\frac{2}{r} \alpha'(r) =0,
 $$
which has a general solution
 $$
 \alpha(r) = \frac{C}{r}+D.
 $$
As $r \to \infty$, $\alpha(r)=0$, hence $D=0$.
 At $r=R$, $\alpha(R)=C/R= {\rm constant}$ which we set it equals to
 $\alpha_0$.
  So, in this region Eq.~(\ref{bubble}) becomes
 $$
 \phi_{\rm bubble}(r,t) = \phi_+ + \alpha_0 \, {\rm sin}\,
 \omega t,
 $$
which is the same equation (\ref{phif}).

Secondly, the region inside the bubble $(r<R)$. Again, in this
case, $\phi_0=\phi_-$ (true vacuum), $\frac{d^2V}{d\phi^2}(\phi_0)
\rightarrow \frac{d^2V}{d\phi^2}(\phi_-) =\omega^2+ k^2$, where
$k^2= 6 g \lambda^4+18 \delta$ and Eq.~(\ref{eom2}) becomes
 $$
 \alpha''(r) +\frac{2}{r} \alpha'(r) - k^2 \alpha(r)=0,
 $$
which has a general solution
 $$
 \alpha(r) = A \frac{{\rm sinh}kr}{kr}.
 $$
 At $r=R$, $\alpha(r)= \alpha_0$ and  $A = \alpha_0
\frac{kR}{{\rm sinh}kR}.$ Hence
 $$
 \alpha(r)=  \alpha_0
\frac{R}{r} \frac{{\rm sinh}kr}{{\rm sinh}kR}.
 $$
Note that when $\alpha(r)=0$, the oscillation decays to zero
inside the bubble. Therefore, the thickness of this region
($\Delta$) is given by
 $$
  \Delta = \frac{1}{\sqrt{6 g
 \lambda^4+18 \delta}} \simeq \frac{1}{\sqrt{18 \delta}}
 $$
for small values of $\delta$. Since
 $$
 kR \sim \frac{12 \sigma_0}{\lambda^2 \sqrt\delta} >> 1,
 $$
then the solution for $\alpha(r)$ can be approximated to
\be
 \alpha(r) = \alpha_0 \frac{R}{r} \frac{{\rm e}^{kr}-{\rm
 e}^{-kr}}{{\rm
e}^{kR}-{\rm e}^{-kR}} \simeq \alpha_0 \frac{R}{r} {\rm
e}^{(r-R)/\Delta}
 \ee

As pointed out in \cite{esko}, there are three scales which
characterizes the structure of the oscillating bubbles: the radius
of the bubble $R \simeq 3\sigma_0/(\lambda^2\delta)$, the
thickness of the bubble wall $L \simeq 1/\mu \simeq
1/\sqrt{\frac{d^2V}{d\phi^2}(\phi_-)} \simeq 1/(\sqrt{8g}
\lambda^2)$ and the thickness of the region inside the bubble
where the oscillations decay $\Delta  \simeq 1/(\sqrt{18 \delta})$
and they are related as $
 L << \Delta << R. $

Finally, the region near the wall $(r \sim R).$ In this case
$\phi_0(r)=\phi_{\rm wall}(r)$ and it is computed when the
potential is degenerate, i.e., when $\delta \rightarrow 0$ and is
satisfying the differential equation
 $$
 \frac{d^2\phi_{\rm wall}(r)}{dr^2}=\frac{dV}{d\phi}({\phi_{\rm
 wall}}),
 $$
which has a solution
 $$
 \phi_{\rm wall}(r) = \frac{\lambda^2}{1+{\rm e}^{\mu r}},
 $$
where $\mu =V''(\phi_-)=\sqrt{8 g} \lambda^2$ up to a correction
of first order in $\delta$ and it is the mass of excitations
around the true vacuum. Since we are working within the frame of
the TWA, we can neglect the term $\frac{2}{r} \alpha'(r)$ in
Eq.~(\ref{eom2}) and we approximate $\omega^2$ to  $\omega^2
\simeq 2 g \lambda^4$. Then Eq.~(\ref{eom2}) becomes
 $$
 \alpha''(r) +\left[\omega^2
-\frac{d^2V}{d\phi^2}(\phi_{\rm wall} \right]
 \alpha(r) =0
 $$
which can be simplified to
 $$
 \alpha''(r)+ 6 g \lambda^4 \left[ \frac{4}{1+{\rm e}^{\mu(r-R)}}
 - \frac{5}{(1+{\rm e}^{\mu(r-R)})^2} \right] \alpha(r)=0
 $$
By assuming $x=\mu(r-R)$, then the above equation becomes

\be
 \alpha''(x)+ \frac{3}{4} \left[ \frac{4}{1+{\rm e}^{x}}
 - \frac{5}{(1+{\rm e}^{x})^2} \right] \alpha(x)=0. \label{alphax0}
\ee
We have solved the above equation numerically which is shown in
Figure 2. One can interpolate the solution to an approximate
function given by
\be
 \alpha(r) \simeq  \frac{B}{4} \Bigg[0.013 (\mu(r-R))^2 + 5.0 \, {\rm
 tanh}^20.15
 (\mu(r-R))
 -1.0  \Bigg] \label{alphax}
\ee
Since $\alpha(r) \to B$ in regions $r\leq R$, where we know that
$\alpha(r)=\alpha_0$, we set $B=\alpha_0$.

\begin{figure}
\epsfig{file=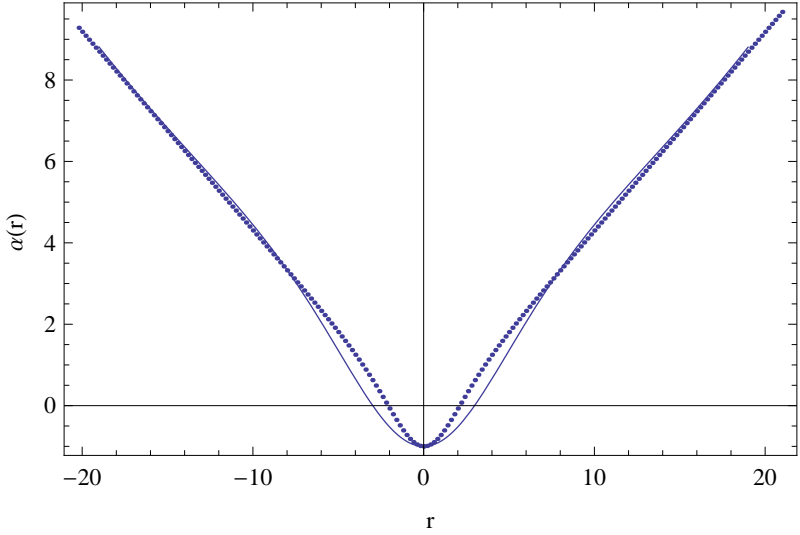}
   \caption{ The dots represents the numerical solution of
Eq.~(\ref{alphax0}) while the solid line as an approximate fit
(Eq.~(\ref{alphax})).}
\end{figure}

To summarize, we have found a solution for the oscillating bubble
in the thin-wall approximation
\be
 \phi_{\rm bubble}(r,t) = \phi_0(r) + \alpha(r) \, {\rm sin}\,
 \omega t,
\ee
where $\phi_0(r)$ is the static solution given by
Eq.~(\ref{phi0}), and $\alpha(r)$ is given by

\be
 \alpha(r) = \left\{
     \begin{array}{lll}
         \alpha_0, &  r \ge R + \frac{L}{2}  \\[0.3cm]
   \frac{\alpha_0}{4} \Big[ 0.013 (\mu(r-R))^2 + 5.0 \,
   {\rm tanh}^20.15\mu(r-R) -1 \Big], &   R-\frac{L}{2} \le r \le
   R+\frac{L}{2}
   \\[0.3cm]
 \alpha_0 \frac{R}{r} {\rm
e}^{(r-R)/\Delta} , & r \le R + \frac{L}{2}
        \end{array}
        \right. \label{alphar}
\ee

\end{section}

%
%

\begin{section}{\bf Bubble Nucleation Decay Rate}

The bubble nucleation rate per unit time per unit volume is given
by
\be
 \Gamma = A\, {\rm e}^{-2 {\rm Im}[ S(t_0)]}.
\ee

We need a time path which shrinks the bubble to zero size. As
an example of a path is
 $$
 R^2 ={\mathcal R_0^2} + (t-t_0)^2
 $$
 which yields to
 $$
  t = t_0+ i \sqrt{\mathcal R_0^2-R^2},~~ {\rm for}~ R<\mathcal
  R_0.
 $$

We divide the action in Eq.~(\ref{action}) into two parts:

\be
 S_1 = - \int_{t_0+i \mathcal R_0}^{t_0} dt \left[ 4 \pi \sigma_E R^2
 \sqrt{1-\dot
 R^2}
 - \frac{4\pi}{3} \rho_E R^3  \right]
\ee
and
 \be
 S_2 = - \int_{t_0+i \mathcal R_0}^{t_0} dt \left[ 4 \pi  R^2 \sqrt{1-\dot
 R^2}
  \int_{R-\frac{L}{2}}^{R+\frac{L}{2}} dr (\dot \phi^2_{\rm bubble} - \dot
  \phi_f^2)-
  \frac{4\pi}{3} R^3 \dot \phi^2_t
  \right]
 \ee

For the first part $S_1$, the calculations proceed as in the
static case as shown in section $2$. The result is:

\be
 {\rm Im} S_1 = \frac{\pi^2}{12} \rho_E \mathcal R_0^4
\ee
where  $\rho_E$ is the energy density and is given by:

\bea
 \rho_E & = & \rho_E^{ \rm FV} - V(\phi_t) \nonumber \\
  & = & \frac{1}{2} [\dot \phi_f(t)]^2+V(\phi_f)-V(\phi_t)
= \rho_0 + \frac{1}{2} \alpha_0^2 \omega^2, \nonumber
 \eea
which is time independent. Note that the oscillation about the
false vacuum increases the energy density and in the limit
$\alpha_0 \to 0$, $\rho_E=\rho_0$ as expected.

The surface tension $\sigma_E$ is given by
 $$
 \sigma_E = \sigma_E^{\rm bubble} - \sigma_E^{\rm FV},
 $$
where $\sigma_E^{\rm bubble}$ is given by
 $$
 \sigma_E^{\rm bubble}= \int_{\rm wall} dr \left[ \frac{1}{2}
  \left(\dot \phi_{\rm bubble}(r,t)\right)^2 +
  \frac{1}{2} \left( \phi'_{\rm bubble}(r,t)\right)^2 + V(\phi_{\rm
  bubble}).
  \right]
 $$
Using
 $$
  \phi_{\rm bubble}(r,t) = \phi_0(r) + \alpha(r) \, {\rm sin}\,
 \omega t =  \phi_0(r) + \beta(r,t)
 $$
and
 $$
 V(\phi_{\rm bubble}) = V(\phi_0)+ \beta(r,t)
 \frac{dV}{d\phi}(\phi_0)+ \frac{1}{2} \beta^2(r,t)
 \frac{d^2V}{d\phi^2}(\phi_0)
 $$
then
\begin{eqnarray}
 \sigma_E^{\rm bubble}& =& \int_{\rm wall} dr  \Bigg[
  \frac{1}{2} \omega^2 \alpha^2(r) {\rm cos}^2\, \omega t +
  \frac{1}{2} \phi^{'2}_0(r)
    + \frac{1}{2} \beta^{'2}(r,t)
   + \phi_0'(r)\beta'(r,t)+ V(\phi_0) \nonumber \\
   & +& \beta(r,t) \frac{dV}{d\phi}(\phi_0) +
  \frac{1}{2} \beta^2(r,t) \frac{d^2V}{d\phi^2}(\phi_0)
  \Bigg]
  \nonumber \\
  & = & \sigma_0 + \sigma_1 + \sigma_2, \nonumber
\end{eqnarray}
where
 $$
 \sigma_0= \int_{\rm wall} dr \Big[ \frac{1}{2} \phi^{'2}_0(r) +
 V(\phi_0) \Big],
 $$
and
  $$
 \sigma_1  =  \int_{\rm wall} dr \Big[\phi_0'(r)\beta'(r,t) +
 \beta(r,t) \frac{dV}{d\phi}(\phi_0) \Big],
  $$
which can be shown equals to zero. While
\bea
 \sigma_2 & = & \int_{\rm wall} dr \Big[-\frac{1}{2} \omega^2
 \alpha^2(r) {\rm sin}^2\, \omega t + \frac{1}{2} \alpha'^{2}(r)
 {\rm sin}^2\, \omega t + \frac{1}{2} \beta^2(r,t)
 \frac{d^2V}{d\phi^2}(\phi_0) + \frac{1}{2} \omega^2 \alpha^2
 \Big] \nonumber \\
 & = & \int_{\rm wall} dr \frac{1}{2}\Bigg[ \Bigg( \alpha^2(r) \Big(
   \frac{d^2V}{d\phi^2}(\phi_0) - \omega^2 \Big) + \alpha'^{2}(r)
   \Bigg) {\rm sin}^2 \omega t + \omega^2 \alpha^2(r) \Bigg]
   \nonumber \\
 & = & \int_{\rm wall} dr \frac{1}{2} \Bigg[\Bigg( \alpha(r)
 \alpha''(r)+ (\alpha'(r))^2
 \Bigg) {\rm sin}\,^2 \omega t + \omega^2 \alpha^2(r)
 \Bigg] \nonumber \\
 & = & \frac{1}{2} \omega^2  \int_{\rm wall} dr \alpha^2(r)
  =  0.06 \frac{\omega^2}{2} L \alpha_0^2 \nonumber
\eea
Therefore,
 $$
 \sigma_E^{\rm bubble} = \sigma_0 +  0.06 \frac{\omega^2}{2}
 \alpha_0^2 L
 $$
while the value surface density due to the false vacuum is
 $$
 \sigma_E^{\rm FV}  =  \int_{\rm wall} dr \Big[\frac{1}{2}
 \dot\phi_f^2(r,t) + V(\phi_f) \Big]
  =
 \frac{1}{2} \alpha_0^2 \omega^2
 \int_{R-\frac{L}{2}}^{R+\frac{L}{2}} dr
  =  \frac{1}{2} \alpha_0^2 \omega^2 L
  $$
Hence,
 \be
 \sigma_E  =  \sigma_0 -  0.47 \omega^2  \alpha_0^2 L
\ee
which is again independent of time, but the oscillation decreases
its value and in the limit $\alpha_0 \to 0$, $\sigma_E=\sigma_0$
as expected. Moreover, notice that $\sigma_E$ equals to zero when
$\alpha_0^2 =\sigma_0/(0.47 \omega^2 L).$ We would like to see the
effect of $\alpha$ on the radius of the bubble. Figure $3$ shows
the ratio of the relative difference of the radius of the bubble
with oscillation about the false vacuum ($\mathcal R$) and radius
without oscillation ($R_0$) versus $\alpha_0$. We notice from the
figure that at $\alpha_0=0$ the value of $\mathcal R$ equals to
$R_0$ and at $\alpha_0=2.57$ its value is zero for $\delta=0.2$.
So, the allowed value of $\alpha_0$ is $0<\alpha_0 < 2.57$.

\begin{figure}
\epsfig{file=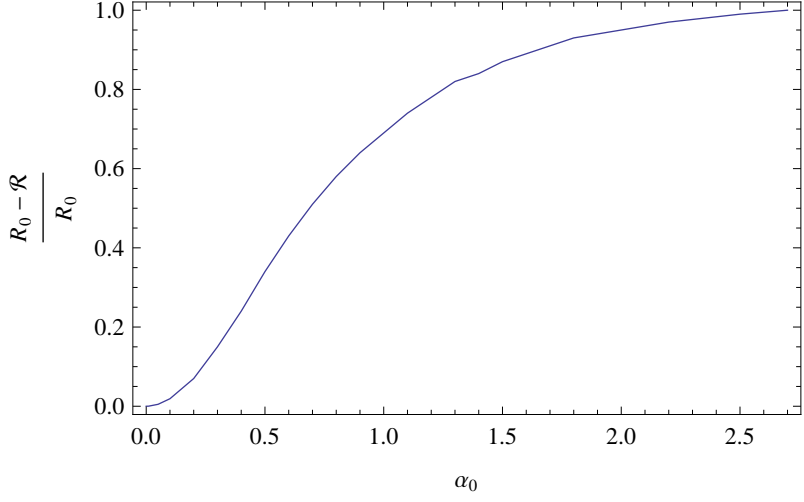}
   \caption{ The ratio of the relative difference of the radius of the bubble
with oscillation about the false vacuum and radius without
oscillation versus $\alpha_0$.}
\end{figure}

The second part of the action $S_2$ is given by
\bea
 S_2& =& - \int_{t_0+i \mathcal R_0}^{t_0} dt \Bigg[ 4 \pi R^2(t) \sqrt{1-\dot
 R^2(t)}
 \int_{\rm wall} dr \Big( \dot\phi_{\rm bubble}^2(r,t)- \dot\phi_f^2(r,t) \Big)
  - \frac{4 \pi}{3} R^3(t) \dot\phi_f^2(r,t)  \Bigg] \nonumber \\
 &=& - 4\pi \omega^2 \alpha_0^2 \int_{t_0+i \mathcal R_0}^{t_0} dt \Bigg[L D
 R^2(t)
 \sqrt{1-\dot
 R(t)^2} - \frac{1}{3} R^3(t)  \Bigg] {\rm cos}^2\omega t
 \nonumber
 \eea
where $D=-0.94$. Using $R(t)=\sqrt{ \mathcal R_0^2+(t-t_0)^2}$ and
$t-t_0 = {\mathcal R_0} z i$, then

\bea
 S_2 &  =& - 4\pi \omega^2 \alpha_0^2
 \Bigg[  \frac{i}{2} {\mathcal R_0} \int_{1}^{0} dz  \left(L D {\mathcal R_0^2}
 \sqrt{1-z^2}
 - \frac{1}{3} {\mathcal R_0^3} (1-z^2)^{\frac{3}{2}} \right) \nonumber \\
 & + & i {\mathcal R_0} \int_{1}^{0} dz  L D {\mathcal R_0^2} \sqrt{1-z^2}
 \frac{1}{4} \Big( {\rm e}^{2i\omega t_0} {\rm e}^{-2\omega {\mathcal R_0} z
} + {\rm e}^{-2i\omega t_0}{\rm e}^{2 \omega {\mathcal R_0} z}
\Big)
\nonumber \\
 & + & i {\mathcal R_0} \int_{1}^{0} dz  \Big(
 - \frac{1}{3} {\mathcal R_0^3} (1-z^2)^{\frac{3}{2}} \Big)
 \frac{1}{4} \Big( {\rm e}^{2i\omega t_0} {\rm e}^{-2\omega { \mathcal R_0} z
} + {\rm e}^{-2i\omega t_0}{\rm e}^{2 \omega { \mathcal R_0} z}
\Big) \Bigg]
\nonumber\\
& = &- 4\pi \omega^2 \alpha_0^2 ( I_1+I_2+I_3) \nonumber
 \eea
It can be easily shown that
 \bea
{  {\rm Im} I_1} & = & \frac{1}{2} { \mathcal R_0} \int_{1}^{0} dz
\Bigg[L D { \mathcal R_0^2} \sqrt{1-z^2}
 - \frac{1}{3} {\mathcal R_0^3} (1-z^2)^{\frac{3}{2}} \Bigg] \nonumber \\
 & = & \frac{\pi}{16 \omega}(0.94) {\mathcal R_0^3} +
 \frac{\pi}{32} \mathcal R_0^4 . \nonumber
\eea
and
\bea
 {\rm Im} I_2 & =&   { \mathcal R_0^3} D L \int_{1}^{0} dz  \sqrt{1-z^2}
 \frac{1}{4} \Big({\rm cos}\, 2\omega t_0 \Big) \Big( {\rm e}^{-2\omega { \mathcal R_0} z
} + {\rm e}^{2 \omega {\mathcal R_0} z} \Big) \nonumber \\
 & = & - {\mathcal R_0^3} D L \frac{1}{4} \Big({\rm cos}\, 2\omega t_0 \Big) \int_{-1}^{1}
 dz
 \sqrt{1-z^2}
  {\rm e}^{2 \omega {\mathcal R_0} z} \nonumber \\
 & = & - \frac{ D }{2\omega} \frac{{\mathcal R_0^3}}{4}
\frac{\pi}{2\omega \mathcal R_0} \Big({\rm cos}\, 2\omega t_0
\Big) I_1(2\omega \mathcal R_0) \nonumber
 \eea
by using Modified Bessel Functions
 $$
 I_{\nu}(z) = \frac{(z/2)^\nu}{\sqrt\pi \Gamma(\nu+1/2)}
 \int_{-1}^1(1-t^2)^{\nu-1/2} {\rm e}^{\pm zt} dt
 $$
and $L\simeq 1/\sqrt{8 g \lambda^2} \simeq 1/(2\omega)$. Similarly
for $I_3$,
\bea
  {\rm Im} I_3 & = & \frac{\mathcal R_0^4}{3} \frac{1}{4} \Big({\rm cos}\,
2\omega t_0 \Big) \int_{-1}^{1} dz
 ({1-z^2})^{3/2}
  {\rm e}^{2 \omega {\mathcal R_0} z} \nonumber
 \\
& = & \frac{1}{16} \pi { \mathcal R_0^2} \frac{1}{\omega^2}
\Big({\rm cos}\, 2\omega t_0 \Big) I_2(2\omega \mathcal R_0).
\nonumber
 \eea
Therefore,
\bea
 {\rm Im} S_2
& = & \pi^2 {\mathcal R_0^2} \alpha_0^2 \Bigg(
\Big(\frac{0.94}{8}\Big)(2\omega \mathcal R_0)
+ \frac{(2\omega \mathcal R_0)^2}{32} \nonumber \\
 &+&
 \Bigg((\frac{0.94}{4}) I_1(2\omega  \mathcal R_0)
+ \frac{1}{4} I_2(2\omega \mathcal R_0)\Bigg) {\rm cos}\, 2\omega
t_0 \Bigg)
 \eea
The total instantaneous bubble nucleation rate is then
\bea \Gamma (t_0) &=& {\rm Exp}\Bigg[ -\frac{\pi^2}{6} \rho_E
\mathcal R_0^4 - \pi^2{ \mathcal R_0^2} \alpha_0^2  \Bigg(
\Big(\frac{3.76}{32}\Big)(2\omega  \mathcal R_0)
+ \Big(\frac{1}{32}\Big)(2\omega  \mathcal R_0)^2 \nonumber \\
 &+&
 \Bigg((\frac{0.94}{4}) I_1(2\omega  \mathcal R_0)
+ \frac{1}{4} I_2(2\omega \mathcal R_0)\Bigg) {\rm cos}\, 2\omega
t_0 \Bigg) \Bigg].
 \eea

By fixing the value of $\delta$ to $0.2$, we have shown that total
action of the instanton (${\rm Im} S$) varies with $\alpha_0$. It
has a maximum value ($S_{\rm max.}$) at $\alpha_0 \approx 0.12$,
if we assume ${\rm cos}\, 2\omega t_0=1$. Figure $4$ shows a plot
of ($(S-S_0)/S_{\rm max}$) versus $\alpha_0$ where $S_0$ is the
action give by Eq.~(\ref{S00}). At $\alpha_0=0$, we have $S=S_0$
while at $\alpha_0 \approx 0.12$, most contribution of the action
comes from the oscillatory part $\omega \mathcal R_0$ when it has its maximum value. For $\alpha_0 > 0.12$, the contribution from the
oscillatory part starts decreasing and the total action converges
to $S_0$ for higher values of $\alpha_0$. So, we conclude that
the effect of oscillation about the false vacuum has a significant
contribution to the tunneling for small specific value of
$\alpha_0$.

\begin{figure}
\epsfig{file=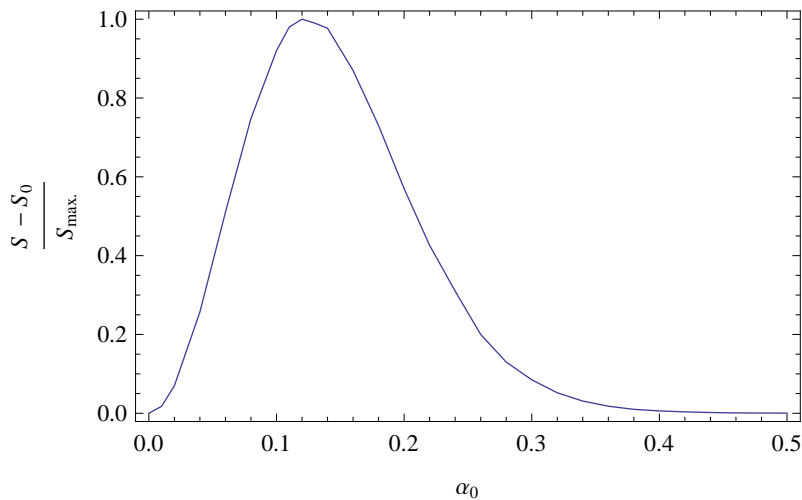}
   \caption{The plot
of ($(S-S_0)/S_{\rm max}$) versus $\alpha_0$. }
\end{figure}

\end{section}
%

%

\begin{section}{\bf Conclusion}

In this paper we have discussed the problem of false vacuum decay
in field theory, where the initial state consists of coherent
field oscillations about the false vacuum for $\phi^6$ potential.
We have shown that there is an upper limit for the amplitude of the
oscillation of the field about the false vacuum. Moreover,  The
effect of oscillation about the false vacuum has a significant
contribution to the tunneling for small specific values of the
amplitude. The method we have used is based on the WKB
approximation and the solutions of classical equation of motion of
the instanton along complex time contour. We obtained a
time-dependent decay rate in the case of small oscillations.

 The importance of our work is for cosmological models which are based
on quantum tunneling, for example: eternal inflation
\cite{vilenkin}, the Hartle-Hakwing-instanton \cite{hartle}, the
Hawking-Moss instanton \cite{hawking}, the quantum creation of
topological defects, e.g. strings and branes in a fixed space-time
\cite{basu}. Moreover, several authors have suggested that string
theory in four dimensions might have many different vacua
\cite{douglas}, which are all represent local minima and the tunneling
between different local minima is of great importance. Finally, we
would like to mention here that an important problem which can be
investigated is the quantum nucleation of cosmic strings and
domain walls in an expanding universe.

\end{section}

\bibliography{plain}


\end{document}